\documentclass{article}
\usepackage{graphicx}
\newcommand{\myscalebox}[1]{\scalebox{0.43}[0.43]{#1}}
\newenvironment{myalgorithm}[1]
{
\begin{tabbing} xx \= xx \= xx \= xx \= xx \= xx \= xx \= xx \kill
 {\bf algorithm #1}\\
 {\bf begin}\\
}
{
 {\bf end}\\
 \end{tabbing}
}

\begin{document}
\title{Statistical mechanics perspective on the phase transition
in vertex covering of finite-connectivity random graphs}
\author{Alexander K. Hartmann and Martin Weigt\\
Institute for Theoretical Physics,\\
University of G\"ottingen, Bunsenstr. 9, 37073 G\"ottingen, Germany\\
E-mail: \texttt{hartmann/weigt@theorie.physik.uni-goettingen.de}
        }

\date{\today}
\maketitle

\begin{abstract}
The vertex-cover problem is studied for random graphs $G_{N,cN}$ having
$N$ vertices and $cN$ edges. Exact numerical results are obtained by
a branch-and-bound algorithm. It is found that a transition in the
coverability at a $c$-dependent threshold $x=x_c(c)$ appears, where
$xN$ is the cardinality of the vertex cover. This transition coincides 
with a sharp peak of the typical numerical effort, which is needed to
decide whether there exists a cover with $xN$ vertices or not. 
Additionally, the transition is visible in a jump of the backbone size
as a function of $x$.

For small edge concentrations $c\ll 0.5$, a cluster expansion is 
performed, giving very accurate results in this regime. These
results are extended using methods developed in statistical 
physics. The so called annealed approximation reproduces a rigorous
bound on $x_c(c)$ which was known previously. The main part of the 
paper contains an application of the replica method. Within the
replica symmetric ansatz the threshold $x_c(c)$ and the critical 
backbone size $b_c(c)$ can be calculated. For $c<e/2$ the results show an 
excellent agreement with the numerical findings. At average vertex 
degree $2c=e$, an instability of the simple replica symmetric solution 
occurs.
\end{abstract}


\section{Introduction}\label{sec:intro}

According to Garey and Johnson \cite{GaJo}, the vertex cover (VC) 
problem belongs to the six basic NP-complete problems. Here VC is
investigated for an ensemble of random graphs $G_{N,cN}$ having $N$
vertices and $cN$ edges \cite{ErRe}, with $c$ constant. 
Despite some efforts in the past \cite{Ga,Fr}, no solution for
the critical cardinality $X_c(c)$ of the vertex cover as a function 
of $c$ has been found, but some lower and upper bounds were obtained. 
In this paper we investigate the problem with an exact branch-and-bound
algorithm, a cluster expansion for small $c$ and with methods borrowed
from the statistical physics of disordered systems \cite{MePaVi}, see
also \cite{WeHa}. 

Our main result is the following, with $W$ being the LambertW-function
($x=W(x)e^{W(x)})$):\\
{\it In the large-N limit and for $c\leq e/2$ ($e$ Eulerian constant),
the cardinality $X_c(c)$ of the minimal vertex cover of a random graph
$G_{N,cN}$ is given by}
\begin{equation}
  \label{schwelle}
  X_c(c) = N - \frac{2W(2c)+W(2c)^2}{4c}N + o(N)\ ,
\end{equation}
{\it and the number of vertices being in the backbone (see below) of these 
minimal VCs reads} 
\begin{equation}
  B_c(c) = N - \frac{W(2c)^2}{2c}N + o(N)\ .
\end{equation}
{\it For $c>e/2$, the expression given on the right-hand side of 
(\ref{schwelle}) provides a lower bound on $x_c(c)$.}

The {\em backbone} is defined as follows: Usually for a graph
different minimal vertex covers exist. A vertex which belongs either to
all vertex covers or to no vertex cover of a given graph is said to
belong to the backbone.

Statistical mechanics methods were already applied to other famous 
NP-complete problems, as {\it e.g.} $K$-satisfiability (KSAT) 
\cite{MoZe} or number partitioning \cite{mertens98}. They are known 
to show interesting phase transitions in their solvability 
and, even more interestingly, in their typical case algorithmic 
complexity, {\it i.e.} in the dependence of the median solution time 
on the system size \cite{review,hayes97}. Consider {\it
e.g.} the satisfiability problem with the number of constraints per
variable as a parameter. When this parameter 
exceeds a certain threshold, the solvability of a randomly
chosen logical formula undergoes a sharp transition from almost always
satisfiable to almost always unsatisfiable \cite{MiSeLe}.  The hardest
to solve formulae are found in the vicinity of the transition point. 
Far away from this point the solution time is much smaller, as the
problem is easily fulfilled or hopelessly over-constrained. The
typical solution times in the under-constrained phase are even found
to depend only polynomially on the system size! Recently, insight
coming from a statistical-physics perspective on these problems
\cite{MoZe} has lead to a fruitful cooperation with computer
scientists, and has shed some light on the nature of this transition
\cite{nature}. Frequently, on the cost of not being mathematically 
rigorous, methods of statistical physics allow to obtain more insight 
than classical tools of computer science or discrete mathematics. 
This is true for the VC problem as well, as will be shown in this 
work.

The paper is organized as follows. After this introductory section, 
the investigated model, related problems, and several notations are 
introduced. Some previously known rigorous bounds for the minimum 
cardinality of the vertex-cover are cited. In the third
chapter VC  is studied numerically with an exact branch-and-bound
procedure. Then a cluster expansion for disconnected graphs 
with low average vertex degree is performed. Section 5 contains the
main part of the paper: statistical physics strategies are applied. 
A short introduction is given, which relates several elements of 
graph theory to corresponding quantities appearing in physics. 
Then, two approaches are presented. The {\em annealed approximation}
 reproduces 
one of the above-mentioned rigorous bounds. More detailed insight is
gained by the {\em replica method}. Using the replica symmetric ansatz, the 
threshold and the backbone size at the threshold can be calculated. The
results are compared with the data obtained by the branch-and-bound 
method. In the last section conclusions and an outlook are given.


\section{The model}\label{sec:model}

\subsection{Vertex cover and related problems}\label{sec:vc}

In this section we want to introduce the investigated model.

Take any graph $G=(V,E)$ with the $N$ vertices $i\in\{1,...,N\}$
and $M$ edges $(i,j)\in E\subset V\times V$. A {\it vertex cover} (VC)
is a subset $V_{VC}\subset V$ of vertices such that for every edge
$(i,j)\in E$ there is at least one of its endpoints $i$ or $j$ in
$V_{VC}$. We call the vertices in $V_{VC}$ covered, whereas the
vertices in its complement $V\setminus V_{VC}$ are called uncovered.

Also {\em partial covers} are considered. In this case the set
$V_{VC}$ is not a VC and there are some edges $(i,j)$ with 
$i\notin V_{VC}$ and $j \notin V_{VC}$. In this case we call the edge
uncovered as well. The task of finding
the minimum number of uncovered edges given a graph $G$ and 
the cardinality $X\equiv |V_{VC}|$ is an optimization problem.

The corresponding {\it decision problem}, whether there exists a
VC $V_{VC}$ of fixed cardinality $X=|V_{VC}|$, with $1\leq X <N$, is
according to Garey and Johnson \cite{GaJo} one of the six basic
NP-complete problems. So it is widely believed that one cannot
construct any algorithm which solves the problem substantially faster
than exhaustive search, {\it i.e.} only algorithms are known which
have an exponential worst-case time complexity in $N$ and $M$.

VC is related to other well-known and widely used NP-complete
problems. The first one is the {\it independent set} (ISET) problem.
An ISET is a subset $V_{ISET}\subset V$ of vertices such that  for
all $i,j\in V_{ISET}$ we have $(i,j)\notin E$. So 
$V\setminus V_{ISET}$ is obviously a VC for every ISET $V_{ISET}$, and 
every maximal ISET is the complement of a minimal VC. The 
{\it independence number}, defined as the maximum of cardinalities 
$|V_{ISET}|$ of all ISETs, is consequently given by 
$N-\min_{\mbox{VC}} |V_{VC}|$.

A {\it clique} is a fully connected subgraph. So, if the subset
$V_{ISET}\subset V$ is an ISET in $G=(V,E)$, it is a clique in the
complementary graph $\overline{G}=(V,V\times V \setminus E)$. Finding
the largest clique in one graph is equivalent to finding the largest ISET in
the complementary graph.

\subsection{Random graphs}\label{sec:rg}

In order to speak of median or average cases, and of phase
transitions, we have to introduce a probability distribution over
graphs. This can be done best by using the concept of {\it random
graphs} as already introduced about 40 years ago by Erd\"os and
R\'enyi \cite{ErRe}.  A random graph $G_{N,M}$ is a graph with $N$
vertices $V=\{1,...,N\}$ and $M$ randomly drawn edges such that any
two instances (for fixed $N,M$) are equiprobable.

An alternative description would be, to include an arbitrary pair
of vertices with a certain probability $p$. For large $N$, the number
of edges becomes almost surely $pN^2/2+O(N)$, and both concepts can 
be identified by choosing $p=2M/N^2$.

The regime we are interested in are {\it finite connectivity graphs}
where the average vertex degree $2c=2M/N$ stays constant in the large
$N$ limit. Under this scaling of the edge number, the cardinality 
of the minimal VC should typically depend linearly on $N$ as well, 
$\min_{\mbox{VC}} |V_{VC}|= x_c(c) N$. The main purpose of this paper
is to show evidence 
that there is an asymptotically ($N\to\infty$) sharp threshold 
$x_c(c)$ which depends for almost all graphs only on the average 
vertex degree $2c$, and to find its functional dependence on c.

Here we want to review shortly some of the fundamental results on
random graphs which were already described in \cite{ErRe}, and which 
are important for the following sections:

The first point we want to mention is the distribution of vertex 
degrees $d$, in the limit $N\to\infty$ it is given by a 
Poisson-distribution with mean $2c$:
\begin{equation}
  \label{formulaPo}
  Po_{2c}(d) = e^{-2c} \frac{(2c)^d}{d!}\ . 
\end{equation}
A second point which is important for the understanding of the following 
is the component structure. For $c<1/2$, {\it i.e.} if the vertices have 
in average less than one neighbor, the graph $G_{N,cN}$ is built up from 
connected components which have up to $O(\log N)$ vertices. The probability
that a component is a specific tree $T_k$ of $k$ vertices is given by
\begin{equation}
  \label{components}
  \rho (k) = e^{-2ck} \frac{(2c)^{k-1}}{k!}\ ,
\end{equation}
and is equal for all $k^{k-2}$ distinct trees. As the fraction of 
vertices which are collected in finite trees is 
$\sum_{k=1}^{\infty} \rho (k) k^{k-2} k = 1$ for all $c<1/2$, in this
case almost all vertices are collected in such trees.
For $c>1/2$ a giant component appears which contains a finite fraction
of all vertices. $c=1/2$ is therefore called the {\em percolation threshold}.

\subsection{Rigorously known bounds}\label{sec:bound}

In this subsection we are going to present some previously known 
rigorous bounds on $x_c(c)$. A general one for arbitrary, 
{\it i.e.} non-random graphs  was given by Harant \cite{Ha}
who generalized an old result of Caro and Wei \cite{CaWe}. Translated 
into our notation, he showed that
\begin{equation}
  \label{bound_harant}
 x_c(G)\leq 1-\frac{1}{N}\frac{\left(\sum_{i\in V}\frac{1}{d_i+1} 
                                   \right)^2}{
                  \sum_{i\in V}\frac{1}{d_i+1} - \sum_{(i,j)\in E} 
                      \frac{(d_i-d_j)^2}{(d_i+1)(d_j+1)}}
\end{equation}
where $d_i$ is the vertex degree of vertex $i$. Using the distribution
(\ref{formulaPo}) of vertex degrees and its generalization to pairs of 
connected vertices, this can easily be converted into an upper bound
on $x_c(c)$ which holds almost surely for $N\to\infty$.

The vertex cover problem or the above-mentioned related problems
were also studied in the case of random graphs, and even completely 
solved in the case of infinite connectivity graphs, where any edge
is drawn with finite probability $p$, such that the expected number of 
edges is $p {N\choose 2}=0(N^2)$. There the minimal VC has cardinality
$(N-2\log_{1/(1-p)}N-O(\log \log N))$ \cite{BoEr}. Bounds in the 
finite-connectivity region of random graphs with $N$ vertices and $cN$
edges were given by Gazmuri \cite{Ga}. He showed that
\begin{equation}
  \label{bound_gazmuri}
  x_l(c) < x_c(c) < 1- \frac{\log 2c}{2c}
\end{equation}
where the lower bound is given by the unique solution of
\begin{equation}
  \label{low}
  0=  x_l(c) \log  x_l(c) + (1- x_l(c)) \log (1- x_l(c))
      - c (1- x_l(c))^2\ .
\end{equation}
As we will see later on, this bound coincides with the so-called
annealed bound in statistical physics. The correct asymptotics for
large $c$ was given by Frieze \cite{Fr}:
\begin{equation}
  \label{asympt}
  x_c(c) = 1 - \frac{1}{c}(\log c - \log\log 2c +1 )+o(\frac{1}{c})\ .
\end{equation}


\section{Numerical evidence for a phase transition}\label{sec:num}

To achieve a thorough insight into the nature of the problem,
numerical simulations were performed. At first the branch-and-bound
algorithm is explained which was implemented for this purpose. Then,
results are presented which relate the transition in solvability to a
change in the median-case time complexity. Also the dependence of the
backbone (see below) on the cover size $x$ shows a jump at this
transition.

\subsection{The algorithm}

All numerical results were obtained by an exact enumeration. Using a
branch-and-bound algorithm similar to
\cite{Lawler66,Tarjan77,Shindo90} all covers can be calculated: as
each vertex is either covered or uncovered, there are $2^N$ possible
configurations which can be arranged as leafs of a binary
(backtracking) tree. At each node, the two subtrees represent the
subproblems where the corresponding vertex is either covered or
uncovered.  The {\em branch} operation tries to find a solution by
investigating both subtrees and keeping only the optimum
solutions.

First we concentrate on the algorithm which finds the configurations
with the minimum number of uncovered edges for a given graph and a
given number $X$ of vertices which can be covered. We want to omit
subtrees which for sure contain no optimum solutions: this is the case
either if the number of
covered vertices exceeds $X$ or if the leafs of the subtree can
already be proven to be worse than previously considered
configurations. Thus, it is possible to avoid branching into some subtrees
by calculating the following {\em bound}: it uses the {\em current}
vertex degree $d(i)$, which is the number of uncovered neighbors at a
specific stage of the calculation. By covering a vertex $i$ the total
number of uncovered edges is reduced by exactly $d(i)$. If several
vertices $j_1,j_2,\ldots,j_k$ are covered, the number of uncovered
edges is {\em at most} reduced by $d(j_1)+d(j_2)+\ldots
+d(j_k)$. Assume that at a certain stage within the backtracking tree,
there are $uncov$ edges uncovered and still $k$ vertices to
cover. Then a lower bound $M$ for the best solution which can be
found in the subtree is
\begin{equation}
M=\max\left[0,uncov-\max_{j_1,\ldots,j_k}d(j_1)+\ldots+d(j_k)\right]\ .
\end{equation}
The maximum is easily calculated by always storing the uncovered
vertices sorted according their current degrees.  The algorithm can
avoid branching into a subtree if $M$ is strictly larger than the
number $opt$ of uncovered edges in the best solution found so far.  If
one is interested only in an arbitrary minimum configuration instead
of enumerating all, one can omit every subtree with $M\geq opt$. In
the latter case the algorithm can be stopped as soon as a
configuration with $opt=0$ is found.

For the order the vertices are selected to be (un-)covered
within the algorithm, the following heuristic is applied:
the order of the vertices is given by their current degree. Thus, the 
first descent into the tree is equivalent to the greedy heuristic which
iteratively covers vertices by always taking the vertex with the
highest current degree. Later, it will be become clear from the
results that this heuristic is indeed a suitable strategy.

The following representation summarizes the algorithm for enumerating
all configurations exhibiting a minimum number of uncovered edges.
Let $G=(V,E)$ be a graph, $k$ the number of vertices to cover and
$uncov$ the number of edges to cover. Initially $k=X$ and $uncov=|E|$.
The variable $opt$ is initialized with $opt=|E|$ and contains the
minimum number of uncovered edges found so far. The value of $opt$ is
passed via call by reference.  At the beginning all vertices $i\in V$
are marked as {\it free}. The marks are considered to be passed via
call by reference as well (not shown explicitly). 
Additionally it is assumed that
somewhere a set of (optimum) solutions can be stored.

\begin{myalgorithm}{min-cover($G,k,uncov,opt$)}
\> {\bf if} k=0 {\bf then} $\{$leaf of tree reached?$\}$\\
\> {\bf begin}\\
\>\> {\bf if} $uncov<opt$ {\bf then }$\{$new minimum found?$\}$\\
\>\> {\bf begin}\\
\>\>\> $opt:=uncov$;\\
\>\>\> clear set of stored configurations;\\
\>\> {\bf end};\\
\>\> store configuration;\\
\> {\bf end};\\
\> {\bf if} bound condition is true (see text) {\bf then}\\
\>\> {\bf return};\\
\> let $i\in V$ a vertex marked as {\it free } of maximal current degree;\\
\> mark $i$ as {\it covered};\\
\> $k:=k-1$;\\
\> adjust degrees of all neighbors $j$ of $i$: $d(j):=d(j)-1$;\\
\> {\bf min-cover}($G,k,uncov-d(i),opt$) $\{$branch into 'left' subtree$\}$;\\
\> mark $i$ as {\it uncovered};\\
\> $k:=k+1$;\\
\> (re)adjust degrees of all neighbors $j$ of $i$: $d(j):=d(j)+1$;\\
\> {\bf min-cover}($G,k,uncov,opt$) $\{$branch into 'right' subtree$\}$;\\
\> mark $i$ as {\it free};\\
\end{myalgorithm}

In the actual implementation, the algorithm does not descend further
into the tree as well, when no uncovered edges are left. In this case
the vertex covers of the corresponding subtree consist of the vertices
covered so far and all possible selections of $k$ vertices among all
uncovered vertices.

Now we discuss the case of finding a true VC of minimum cardinality,
where the performance of the method can be enhanced by some
extensions.  The algorithm is called with $k=|V|, opt=0$ and $k$ is
passed via call by reference like $opt$.  Now assume that during the
execution of the algorithm a total cover ($uncov=0$) is found and
$k>0$. Thus it is possible to cover all edges with less than the
allowed number of vertices.  Consequently, it is not necessary to
cover additional vertices, and the value of $k$ is set to
zero. Additionally the set of configurations which was stored before
is cleared.  Furthermore, whenever a vertex $i$ is marked as {\it
uncovered}, all its neighbors $j$ can be covered immediately, because
no uncovered edge should remain. Please note that in this case the
degrees of all neighbors of the neighbors $j$ of $i$ have to be
readjusted as well. After the initial call of this modified algorithm
has finished, the variable $k$ contains the cardinality of the minimum
vertex cover.

The algorithm was implemented via the help of the LEDA library
\cite{leda99} which offers many useful data types and algorithms for
linear algebra and graph problems. Since the VC problem is NP-hard,
the method exhibits an exponential worst-case time complexity.
Although our algorithm is very simple, in the regime $0.5<c<5$ random
graphs up to size $N=100$ could be treated for all values
$X\in[0,N]$. For the calculation of covers of minimum cardinality,
also graphs with $N=140$ could be considered.  Please note that for
$c<0.5$ the graphs can be divided into many connected components of
sizes up to $O(\log N)$. Then, in the case one is interested only in
the cover of minimum cardinality, the algorithm can be applied to each
component separately, yielding only a polynomial time-complexity.

\subsection{Numerical results}

A first evidence for a peak of the typical case complexity near the
threshold was given in \cite{GeWa} where the problem was matched to
SAT and solved with the Davis Putnam procedure. The running time was
measured for graphs of size $N=12$. Here, systems up to size $N=140$
are investigated. Since data for several different graph sizes are
available, it is possible to extrapolate the behavior of the infinite
graph using finite-size scaling techniques. The results of this
extrapolations will be presented in a subsequent chapter, along with
the outcomes of analytical calculations.

In Fig. \ref{figPEX} the probability $P_{cov}(x)$ of finding a vertex
cover of cardinality $xN$ for a random graph $G_{N,cN}$ is displayed
for $c=1$ and different values of $N$ (10000 instances per value of
$x$, 1000 for $N=100$). The drop of the probability from one for large
cover sets to zero for small cover sets obviously sharpens with
$N$. Thus, a jump at a well-defined $x_c(c)$ is to be expected in the
large-$N$ limit: Above $x_c(c)$ almost all random graphs with $cN$
edges are coverable with $xN$ vertices, below $x_c(c)$ almost no graph
has such a VC.  The curves in the left part of the figure show the
average minimal fraction $e(x)$ of uncovered edges, which for a
coverable graph is obviously zero.  In the large-$N$ limit, the
disappearance of positive $e(x)$ coincides with the threshold.

It is very instructive to measure the median computational effort, as
given by the number of visited nodes in the backtracking tree, in
dependence on $x$ and $N$.  The curves which are exposed in
Fig. \ref{figTimeX}, show a pronounced peak at the threshold
value. Inside the coverable phase, $x>x_c(c)$, the computational cost
is growing only linearly with $N$, and in many cases the heuristic is
already able to find a cover with $xN$ vertices. Below the threshold,
$x < x_c(c)$, it is clearly exponential in $N$ (see inset).  This
easy-hard transition resembles very much the typical-case complexity
pattern of 3SAT \cite{nature}, and deserves some more detailed
investigation, which will be provided by the analytical calculation
later on.

In Fig. \ref{figTimeCovUncov} the median time is plotted separately
for the subset of coverable and uncoverable graphs, respectively. In
addition, a scatter plot is included, which contains a dot for each
result for 100 graphs and for different cardinalities $xN$ of the
cover.  For a given graph $G_{N,cN}$, as long as it is not coverable
with $xN$ vertices, the computer time grows heavily with $x$. But as
soon as a graph is coverable, it takes only a small computational
effort to find a cover. The reason that 
median effort over all graphs is reduced
for $x>x_c$ is that the fraction of uncoverable graphs decreases
rapidly.

Another quantity is directly related to the transition: The outcome of
the algorithm is a configuration, {\it i.e.}  a vector of marks
telling whether a given vertex is covered or not. For a given graph
and a given fraction $x$ usually different configurations are
feasible, exhibiting all the same minimal number $e(x)cN$ of uncovered
edges. An enumeration shows that the number of these configurations
grows exponentially with the system size for all values of
$N$. Nevertheless, for $x<x_c(c)$ there is always a finite fraction of
vertices which behave equally in all different configurations: they
are either always covered or always uncovered. The set of these vertices is
the backbone $B$.

For $x>x_c$ and in the large-$N$ limit, there is no non-empty backbone:
the graph is already coverable with $x_c(G_{N,cN})N$ vertices, the
other $(x-x_c)N$ can be distributed freely. This already excludes the
existence of vertices being always uncovered. The maximal vertex
degree in a random graph $G_{N,cN}$ grows only as $O(\log N)$. So the
neighbors of every covered vertex can be covered with some of the
remaining $(x-x_c)N$ free cover marks, and the central vertex itself
can be uncovered and thus does not belong to the backbone.

Later we will see that directly at the threshold $x=x_c$ a finite
backbone size $b(x)=|B|/N$ appears.  Thus, for $N\to\infty$ the
function $b(x)$ exhibits a discontinuity at $x_c(c)$.  This is
indicated by the results obtained from the numerical calculations,
again for the case $c=1$, see Fig. \ref{figBX}. For $x<x_c(1)$ the
relative backbone size $b(x)$ is large and almost independent of
N. For $x>x_c(1)$ a sharp decrease can be observed, which pronounces
with increasing $N$. A surprising result is obtained, when we study
coverable and uncoverable graphs separately. This can be done only in
the vicinity of the transition, $x\approx x_c(1)$, where coexisting
coverable and uncoverable graphs can be found for finite $N$. The
inset of Fig. \ref{figBX} shows the result: Above the threshold, the
coverable graphs exhibit a smaller backbone, as expected from the
discussion above. But the curves intersect near $x_c(1)$. This
behavior is observed for all graph sizes $N$, and the effect becomes
more pronounced with increasing system size. As an explanation, we
take a look at graphs being coverable with a small number of
vertices. Their distribution of vertex degrees must deviate
substantially from (\ref{formulaPo}), showing more vertices with high
degree. These vertices are expected to be in the backbone with high
probability, see also the discussion 
on the correlation between vertex degree and backbone 
at the end of section \ref{sec:xc}. Consequently, the
backbone is expected to be very large.  The crossing of both curves
close to $x_c$ seems to be accidental.  By measuring the intersection
as a function of $N$ and extrapolating to $N\to\infty$, the limiting
value is found to be significantly below $x_c$.

We have seen that the vertex-cover problem exhibits several peculiar
features. These are worth to be addressed by analytical methods which
allow to reveal the structure of VCs.


\section{Cluster expansion for low vertex degrees}\label{sec:lowc}

One of the classical results on random graphs is, as mentioned in
section \ref{sec:rg}, that for low edge densities $c<1/2$ almost all
vertices are collected in finite trees, as
\begin{equation}
  1 = \sum_{k=1}^{\infty} \rho(k) k^{k-2} k
\end{equation}
with $\rho(k)$ being the distribution of trees $T_k$ with $k$ vertices,
cf. section \ref{sec:rg}. 
So the threshold $x_c$ and the corresponding backbone $b$ are given by
\begin{eqnarray}
  \label{low_c_expansion}
  x_c(c) &=& \sum_{k=1}^{\infty} \rho(k) \left[ \sum_{T_k} 
             X_c(T_k)\right] \nonumber\\
  b(c,x_c(c)) &=& \sum_{k=1}^{\infty} \rho(k) \left[ \sum_{T_k} 
             B_c(T_k) \right]
\end{eqnarray}
where $\sum_{T_k}$ denotes the sum over all different trees $T_k$.
$X_c(T_k)$ (resp. $B_c(T_k)$) is the cardinality of the minimal VCs
(resp. of their backbone) of $T_k$.

For very small average vertex degrees $c\ll 0.5$ the most vertices are
furthermore concentrated in small components, and we can produce good
approximations for the threshold, the backbone etc. by counting small
trees. There also the distinction between backbone and non-backbone
vertices becomes evident: Consider e.g. a connected component
consisting only of two vertices and one edge. To cover this minimally,
we need exactly one vertex -- but it is not specified which one. The
vertices do not belong to the backbone at threshold, and they give a
contribution to a finite entropy ({\it i.e.} an exponential number) of
minimal VCs.  The situation is different for a tree of three vertices
and two edges. The minimal cover is unique: Only the central vertex
has to be taken. Consequently all these three vertices belong to the
backbone at the threshold. Already at this point, the partial freezing
of degrees of freedom as observed in SAT \cite{MoZe,nature} becomes
evident.

We have counted the optimal covers for trees up to 7 vertices, see the
results in table 1. The values for the threshold and the
backbone are lower bounds as a certain fraction of vertices is not
included. Upper bounds are provided by adding the fraction of missing
vertices to the lower bounds. For small $c$ these bound are very
precise, {\it e.g.} for $c=0.1$, $99.98\%$ of all vertices are already
included in the small trees up to size 7. These approximate values
will be a useful testing ground for the statistical mechanics
calculations which are given in section \ref{sec:sm}.

This tree size expansion is not longer possible above the percolation
threshold $c=1/2$. There the giant component arises which includes a
finite fraction of all vertices.

\begin{table}[bt]
{
$$
\begin{array}{||c|c|c|c|c|c|c|c|c||}\hline \hline 
c &0.05 &0.1 &0.15 &0.2 &0.25 &0.3 &0.4 &0.5 \\ \hline\hline
\nu &0.999997&0.9998&0.998 &0.991 &0.97 &0.94 &0.84 &0.71 \\ \hline 
x_{min} &0.045576&0.0840&0.116&0.143&0.16&0.17&0.17&0.15\\ \hline
x_{max} &0.045579&0.0842&0.118&0.151&0.19&0.23&0.33&0.44\\ \hline
b_{min} &0.916684&0.8572&0.812&0.774&0.74&0.70&0.61&0.51 \\ \hline 
b_{max} &0.916687&0.8574&0.814&0.781&0.77&0.76&0.77&0.80 \\ \hline 
s_{min} &0.028774&0.0488&0.063&0.073&0.078&0.08&0.08&0.07\\ \hline 
s_{max} &0.028775&0.0489&0.064&0.076&0.088&0.10&0.13&0.17\\ \hline \hline 
x_c(c) &0.045577&0.0841&0.117&0.146&0.173&0.196&0.237&0.272\\ \hline
b_c(c) &0.916686&0.8573&0.813 &0.779 &0.753 &0.731 &0.700 &0.678 \\ 
\hline\hline 
\end{array}
$$
}

\caption{Results of the cluster expansion for trees having up to 7
vertices and several values of $c$. $\nu$ denotes the fraction of
vertices which are included in the considered trees, $Nx_{min/max}$
give lower and upper bounds on the number of vertices which are
needed to cover these components, $b_{min/max}$ are backbone bounds,
$s_{min/max}$ bounds for the VC entropy. These values are to be
compared with the analytical results of the replica approach which
are presented in the last two lines.}
\label{tab:expansion}
\end{table}


\section{Statistical mechanics approach}\label{sec:sm}

In this section we use the strong similarities between combinatorial
optimization and statistical mechanics. The cost function of a system
which shall be optimized corresponds to the {\it energy function} (or
{\it Hamiltonian}) in statistical mechanics. The elements of the
definition space of the cost function are called microscopic {\it
configurations}. The main aim of statistical mechanics is the
description of the macroscopic behavior of a microscopically defined
model, {\it e.g.} the prediction and description of phase transitions.

\subsection{General strategy}\label{sec:f}

In order to describe the VC phase transition also beyond the
percolation threshold, we are going to use the tools of the
statistical mechanics of disordered systems \cite{MePaVi}. We
therefore map the random graph to a disordered spin system with an
Hamiltonian which shall be minimized. A canonical choice for the
``energy'' of a subset $\tilde{V}\subset V$ of vertices is given
by the number of uncovered edges:
\begin{equation}
  \label{hamiltonian}
  H(\{S_i\},\{J_{i,j}\}) = \frac{1}{2} \sum_{i,j=1}^N J_{i,j} 
        \delta_{S_i,-1} \delta_{S_j,-1}
\end{equation}
where $J_{i,j}$ are the entries of the symmetric adjacency matrix,
they are equal to one whenever there is an edge connecting the
vertices $i$ and $j$, and zero else. The diagonal elements are
identically set to zero. The covering state of the vertices is mapped
to a configuration of $N$ Ising-spins $S_i=\pm 1$: we choose $S_i=+1$
if $i\in \tilde{V}$, {\it i.e.} if the vertex $i$ is covered, and
$S_i=-1$ if $i$ is uncovered. Non-zero contributions to the Hamiltonian
result only from edges having two uncovered endpoints.

The decision problem whether there exists any VC with $xN$ vertices
can be answered by minimizing $H$ under the constraint
\begin{equation}
  \label{constraint}
  \frac{1}{N}\sum_{i=1}^N S_i = 2x-1
\end{equation}
which fixes the cardinality of the cover set, or in physical terms,
the global magnetization of our Ising-spin system. If this restricted
minimal energy equals zero, then there are no uncovered edges left,
and the decision problem can be positively answered. If, on the other
hand, a positive minimal energy is found, there does not exist any VC
of cardinality $xN$, but the ground state energy gives the best
compromise by describing the configuration with the minimal number of
uncovered edges.

In statistical mechanics every microscopic configuration 
$\{S_i\}_{i=1,..,N}$ is assigned a probability proportional to the 
Gibbs-weight $\exp\{ -T^{-1}H(\{S_i\}) \}$ at temperature $T$.
By decreasing $T$, this weight becomes more and more concentrated
in low-energy configurations and finally, at $T=0$, counts only the
{\it ground states}, {\it i.e.} the configurations minimizing the
Hamiltonian. In order to characterize these in the VC problem, 
we introduce at first a {\it non-zero formal temperature}
$T$ and calculate the {\it partition function}
\begin{equation}
  \label{partition}
  Z(T,x|\{J_{i,j}\}) = \sum_{{\mathcal{C}}_x(\{S_i\})}  
    \exp\left\{ -\frac{
    H(\{S_i\},\{J_{i,j}\})}{T} \right\}
\end{equation}
where we sum only over the set ${\mathcal{C}}_x(\{S_i\})$ of
configurations $\{S_i\}_{i=1,..,N}$ which satisfy the magnetization 
constraint (\ref{constraint}). From this we may calculate the {\it
  free-energy density}
\begin{equation}
  \label{f}
  f(T,x|\{J_{i,j}\}) = - \frac{T}{N} \log  Z(T,x|\{J_{i,j}\}) 
\end{equation}
which in its zero temperature limit gives the desired {\it ground
  state energy density}:
\begin{equation}
  \label{e_gs_fintesize}
  e_{GS}(x|\{J_{i,j}\}) = \lim_{T\to 0} f(T,x|\{J_{i,j}\})\ .
\end{equation}
This energy does still depend on the particular realization of the
graph encoded in the matrix $\{J_{i,j}\}$. In the limit $N\to\infty$
(with $c=M/N=const.$) we expect however the free energy to be {\it
  self-averaging }, and so we are only interested in calculating
\begin{equation}
  \label{e_gs}
  e_{GS}(x,c) = \lim_{T\to 0} f(T,x,c)=\lim_{T\to 0}\lim_{N\to\infty}
     \overline{f(T,x|\{J_{i,j}\})}
\end{equation}
where the over bar stands for the average over the ensemble of random
graphs with $N$ vertices and $cN$ edges. Another interesting quantity
is the {\it ground state entropy}
\begin{equation}
  \label{s_gs}
  s_{GS}(x,c) = \lim_{N\to\infty} \frac{1}{N} \overline{\log
    {\mathcal{N}}_{GS}(x,\{J_{i,j}\})  }
\end{equation}
where ${\mathcal{N}}_{GS}(x,\{J_{i,j}\})$ is the number of ground states
with cardinality $xN$ in the graph given by $\{J_{i,j}\}$. It is also
useful to consider the {\it VC entropy}
\begin{equation}
  \label{s_e0}
  s_{VC}(x,c) = \left\{
  \begin{array}{ll} s_{GS}(x,c) & \mbox{if } e_{GS}(x,c)=0\\
                    -\infty     & \mbox{else}
  \end{array}  \right.
\end{equation}
which measures the number of VCs.

\subsection{The annealed approximation}\label{sec:annealed}

Before trying to calculate this, we will present the so-called 
{\it annealed approximation}. We use the bound
\begin{equation}
  \label{ann}
  \overline{\log Z(T,x|\{J_{i,j}\})} \leq \log
  \overline{Z(T,x|\{J_{i,j}\})} 
\end{equation}
for the average of the logarithm of the partition function in terms of
the logarithm of the average of the partition function. It holds because 
the logarithm is a concave function. We easily calculate the annealed 
entropy, see Appendix \ref{app:annealed} for details,
\begin{eqnarray}
  \label{e_ann}
  s_{ann}(x,c) &=& \lim_{T\to 0}\lim_{N\to\infty} \frac{1}{N}
     \log \overline{Z(T,x|\{J_{i,j}\})}\nonumber\\
 &=& -x \log x - (1-x) \log (1-x) - c (1-x)^2
\end{eqnarray}
and can bound the VC entropy
\begin{equation}
  \label{bound}
  s_{VC}(x,c) \leq s_{ann}(x,c)\ .
\end{equation} 
VCs can thus only exist if the annealed entropy is non-negative, and
$x_c(c)$ is bounded from below by $x_{ann}(c)$ which is given by 
$s_{ann}(x_{ann}(c),c)=0$, {\it i.e.} by the inversion of
\begin{equation}
  \label{ann_bound}
  c = \frac{-x_{ann}(c) \log x_{ann}(c) - (1-x_{ann}(c)) \log
    (1-x_{ann}(c))}{(1-x_{ann}(c))^2} \ .
\end{equation}
This is exactly the lower bound given in \cite{Ga} which is not
surprising as Gazmuri used a very similar reasoning.

\subsection{The replica approach}\label{sec:rs}

If we want to go beyond the annealed approximation, we have to average
the logarithm of the partition function over the disorder. 
Unfortunately this cannot be achieved directly, the way out is given
by the so-called {\it replica trick}, a non-rigorous method which
is well-established in the physics of disordered systems
\cite{MePaVi}. Details of the calculation are exposed in appendix
\ref{app:replica}. There we show the derivation of the so-called {\it
  replica symmetric approximation} of the free-energy density
\begin{eqnarray}
  \label{betaf}
f(T,x,c) &=& T\int_{-\infty}^{\infty} 
  \frac{dh\ dk}{2\pi} e^{-ihk} P_{FT}(k) [\log 
  P_{FT}(k)-1] \log 2\cosh T^{-1}h \\ 
&& -cT\int_{-\infty}^{\infty}  
  dh_1 dh_2 P(h_1) P(h_2)\log\left[ 1-(e^{-T^{-1}}-1)
  \frac{e^{-T^{-1}(h_1+h_2)}}{4\cosh T^{-1} h_1\ \cosh T^{-1} h_2}
  \right]\ .\nonumber
\end{eqnarray}
This quantity has to be optimized with respect to the order parameter 
$P(h)$ which is again restricted by the magnetization constraint to
\begin{equation}
  \label{constraint_P}
  2x-1 = \int_{-\infty}^{\infty} dh P(h) \tanh T^{-1} h\ .
\end{equation}
$P_{FT}(k)$ denotes the Fourier-transform of $P(h)$.

The physical interpretation of the order parameter in terms of the
{\it effective field distribution} is straightforward: $P(h)dh$
gives the probability, that a randomly chosen site $i\in V$ has local
magnetization $m_i=\langle S_i \rangle_T = \tanh T^{-1} h$. This
distribution (or the distribution of local magnetizations) is the 
typical order parameter in disordered finite connectivity models,
cf. \cite{KaSo,Mo}. It is determined by the optimization equation for
the free energy (\ref{betaf}) which reads
\begin{equation}
  \label{saddle}
  \int dh\ P(h)\ e^{T^{-1}hs} = \exp\left\{ -2c-\lambda s +2c \int dh\
    P(h)\  \left[ 1+ (e^{-T^{-1}}-1)\frac{1}{1+e^{2T^{-1}h}}
   \right]^{-\frac{s}{2}}  \right\}\ .
\end{equation}
The Lagrange parameter $\lambda$ in the exponential has to be adjusted
in order to meet the magnetization constraint (\ref{constraint_P}).

This equation as well as the expression (\ref{betaf}) for the free
energy still depend on the formal temperature $T$, and the limit 
$T\to 0$ is not totally obvious: we have to
clarify the scaling of the effective fields $h$ with $T$. There are two
main possibilities:
\begin{itemize}
\item The fields $h$ are proportional to the formal temperature,
  $h=O(T)$ for $T\to 0$. As can be simply seen in the expression
  (\ref{betaf}) for the average free energy, we then also have
  $f(T,x,c)=O(T)$, and the ground state energy $e_{GS}(x,c)$
  vanishes. These fields are consequently found in the coverable phase
  with $x>x_c$. Another important property is that the corresponding
  local magnetizations $m=\tanh T^{-1} h$ do not tend to $\pm 1$, and
  the corresponding spins take different orientations in different
  ground states.
\item The fields $h$ remain different from 0 even if the temperature 
  vanishes,
  $h=O(T^0)$. The corresponding spins have $\pm 1$-ground state
  magnetization, and consequently take on the same value in (almost)
  all ground state configurations, {\it i.e.} they form the
  backbone. If we introduce such fields in (\ref{betaf}) we
  immediately find that $f(T,x,c)$ does not vanish in the
  zero-temperature limit, the ground state energy becomes positive, 
  and such fields cannot exist in the COV phase. Their appearance
  marks the transition.
\end{itemize}

\subsubsection{At the threshold}\label{sec:xc}

If we would be able to solve (\ref{saddle}) at finite temperature for
arbitrary $x$ and $c$, we could deduce the scaling directly from the
solution -- and thus we could determine $x_c(c)$.  As this is to
complicated to be achieved directly, we can plug in the two different
scalings, and calculate the limit $T\to 0$. We then find two different 
equations for $P(h)$ in the two different phases. The phase transition 
point is given by the matching of both equations:
\begin{itemize}
\item If we reach the threshold from above, $x\to x_c(c)+0$, we are in
  the coverable phase. According to the above discussion, the
  effective fields are $h = T H_{cov}(x) z$ where $H_{cov}(x)$
  describes the typical absolute value of the field and $z$ is a
  random variable of finite mean and variance. For $x\to x_c(c)+0$ the
  spins $S_i$ are more and more constraint, and at $x_c(c)$ a freezing
  takes place. The limit is therefore described by 
  $H_{cov}(x\to x_c(c)+0)\to \infty$.
\item If we reach the threshold from below, $x\to x_c(c)-0$, we are in
  the uncoverable phase, and at least a finite fraction of all spins
  has to be frozen. The corresponding effective fields scale as 
  $h = H_{uncov}(x) z$ where now the scale for the absolute value of
  $h$ is described by $H_{uncov}(x)$. As we approach the threshold,
  the freezing gets less strong, and $H_{uncov}(x\to x_c(c)+0)\to 0$.
\end{itemize}
In both limits we find the same equation for the probability
distribution $\tilde{P}(z)$ of the rescaled variable $z$,
see appendix B for a derivation. $\mu$ is the
appropriately rescaled Lagrange parameter, it is negative as it
describes a field which decreases the global magnetization from the
maximum entropy point towards the threshold $x_c(c)$:
\begin{eqnarray}
  \label{threshold}
  \tilde{P}(z+\mu) &=& \sum_{d=0}^\infty e^{-2c}\frac{(2c)^d}{d!} 
  \left[\tilde{P_-}^{\star d}\right](z)\nonumber\\
  \tilde{P_-}(z) &=& \Theta(z) \tilde{P}(-z) + \delta(z)
  \int_{-0}^{\infty} dz\ \tilde{P}(z)
\end{eqnarray}
with the Heaviside step function $\Theta(z)$ and the Dirac
distribution $\delta(z)$. $\tilde{P_-}^{\star d}$ denotes
a $d$-fold convolution product. The interpretation of this equation 
is simple: the effective field for a randomly chosen vertex $i$ is
given by the linear superposition of the local field induced by the
Lagrangian multiplier, and the contribution of its $d_i$ neighbors. If
a neighbor has a negative field, then it is uncovered, and thus forces
a positive field on $i$. If it has a non-negative field it does not
imply any non-vanishing field on $i$. As $\tilde{P}(z)$ is the
histogram of fields for all vertices, equation (\ref{threshold}) 
includes the average over the Poisson distribution (\ref{formulaPo}) 
of vertex degrees.

This equation has a very simple solution,
\begin{equation}
  \label{one-peak}
  \tilde{P}(z) = \sum_{m=-1}^\infty
  \frac{W(2c)^{m+2}}{2c(m+1)!} \delta (z+m\mu)\ , 
\end{equation}
with the Lambert-W function $W$ which is simply defined by
\begin{equation}
  \label{lambertw}
  y=W(x) \leftrightarrow x=y\ e^y\ .
\end{equation}
Non-zero fields correspond to frozen (or backbone) spins, whereas the
Dirac peak in $z=0$ describes all spins which flip from one minimal
VC to a next. The backbone size is consequently given by the total
weight of all nonzero fields. From this we can calculate the threshold
and the backbone,
\begin{eqnarray}
  \label{solution}
  x_c(c) &=& 1 - \frac{2W(2c)+W(2c)^2}{4c}\nonumber\\
  b_c(c) &=& 1 - \frac{W(2c)^2}{2c}\ .
\end{eqnarray}
This result is completely consistent with the bounds of section
\ref{sec:lowc} which is particularly interesting for very small $c$
where these bounds are very close, see table 1.  The result for
$x_c(c)$ is displayed in Fig. \ref{figXCC} along with numerical data,
which were obtained by the variant of the branch-and-bound algorithm
which always looks for a cover of minimum cardinality. For each
treated concentration $c$ of the edges and system sizes $N=12$, 17,
25, 35, 50, 70, 100, 140 for 10000 different realizations of the
random graphs (only 1000 for the $n\ge 100$) the threshold was
calculated. The average value is denoted with $x_c(c, N)$. Then for
each value of $c$ the behavior of the infinite graph was extrapolated
by performing a fit of the function $x_c(N)=x_c +aN^{-b}$ to the data,
where $x_c,a$ and $b$ are tunable parameters. The inset shows an
example of such a kind of extrapolation. The result of $x_c$ as a
function of $c$ shows a very good coincidence with the analytic
result. This is true not only for small concentrations but also for a
region beyond the percolation threshold, whereas systematic deviations
appear for larger $c$.

Are there more complicated solutions to (\ref{threshold}) which 
coincide with the numerics also for larger $c$? At first we
remark that this equation is closed under
\begin{equation}
  \label{more_peaks}
  \tilde{P}^{(l)}(z) = \sum_{m=-l}^\infty a_{m}^{(l)}
  \delta\left(z-m\frac{\mu}{l}\right) 
\end{equation}
for every positive integer $l$. The equations for
$a_{-l}^{(l)},...,a_{-1}^{(l)}$ close, all other weights with
non-negative indices follow. A simple analysis of these equations
shows, that for $c<e/2$ they have no non-trivial solution with only 
non-negative weights, up to this point (\ref{one-peak}) gives the only
valuable solution. For $c>e/2$ non-trivial solutions with an arbitrary
number of peaks appear. 

Together with the above mentioned accordance of bounds and numerical
data for low vertex degrees, this leads to the following {\bf
conjecture}: {\it For random graphs with $c\leq e/2$ the exact values
for the covering threshold and the backbone at this threshold are
given by equation (\ref{solution}). For $c>e/2$, the above value for
$x_c(c)$ still gives a lower bound.}

The last statement follows from the fact that in the replica approach
the saddle point with the largest free energy has to be taken. Imagine
now two different values for $x_c$ would be predicted by two different
saddle points. In between these thresholds, one solution already
predicts a positive energy and hence a larger free energy than the
other. This saddle point has to be preferred, and it corresponds to
the larger threshold.

The transition at $c=e/2$ is not yet understood as also the multi-peak
solutions (\ref{more_peaks}) do not coincide with numerical data.
This can be seen in particular from the behavior of the backbone size
-- which is largely overestimated analytically, see
Fig. \ref{figBCC}. Especially the minimum of $b_c(c)$ at $c=e/2$
cannot be found in the numerical data.  The numerical results were
obtained from the enumerating of all possible covers at the threshold
for the same range of concentrations and sizes mentioned above. Also
the same extrapolation technique to obtain the values for the infinite
random graph was applied.

For the discrepancy of the numerical backbone size with the analytical data for
$c>e/2$, there are two possible explanations:
\begin{itemize}
\item In the analytics we count every spin as backbone which has
magnetization tending to $\pm 1$ in the thermodynamic limit, whereas
the in numerics we count only vertices which have magnetization equal
to $\pm 1$ even for finite size. This difference can be rather
drastic: {\it e.g.} for the fully connected graph of $N$ vertices one
needs $N-1$ for a VC, the average magnetization is therefore
$1-2/N$. The analytics would count a backbone one, whereas the strict
backbone vanishes.
\item Above $c=e/2$ (or even above $c=1/2$) replica symmetry breaking
could appear. This would correspond to a clustering of the VCs in
configurations space, cf. \cite{BiMoWe} for a discussion of this
phenomenon for SAT. As was seen there, the backbone size sensitively
depends on this question. This point is still under investigation.
\end{itemize}

Let us go back to $0<c<e/2$ where (\ref{one-peak}) was conjectured to
be exact, and let us extract more information about the minimal VCs
from our solution. Due to the simple geometrical nature of the
underlying graphs, the VC problem allows a much more intuitive way of
understanding results, in contrast for example to SAT. A first example
was already given in section \ref{sec:lowc} where we gave simple
examples for backbone and non-backbone structures. Let us now
investigate the influence of the close environment of a vertex on its
behavior, more precisely the influence of the vertex degree. The
total distribution of (almost all) degrees is given by the Poisson law
(\ref{formulaPo}), but we can distinguish three distinct
contributions:
\begin{itemize}
\item The joint probability $P(d,m=-1)$ that a vertex has degree $d$ 
and magnetization $m=-1$, {\it i.e.} this vertex belongs to the 
backbone and is uncovered in all minimal VCs.
\item $P(d,m=+1)$ gives the probability that a vertex has degree $d$
and is covered in all minimal VCs.
\item The remaining part of vertices are not in the backbone, thus
described by $P(d,-1<m<+1)$.
\end{itemize}
These quantities can be easily computed from $\tilde{P}(z)$: according
to the interpretation of the self-consistent equation (\ref{threshold})
we can calculate the effective-field distribution for a vertex
of degree $d$ which, in average, has typical neighbors:
\begin{equation}
  \label{P_d}
  \tilde{P}_d (z+\mu) =  \left[\tilde{P_-}^{\star d}\right](z)
\end{equation}
where $\tilde{P_-}(z)$ is exactly the quantity given in 
(\ref{threshold}). Plugging our solution (\ref{one-peak}) into
this equation, we find
\begin{eqnarray}
P(d,m=-1) = \tilde{P}_d (z<0) Po_{2c}(d) &=&
    e^{-2c} \frac{[2c-W(2c)]^d}{d!} \nonumber\\
P(d,-1<m<+1) = \tilde{P}_d (z=0) Po_{2c}(d) &=&
    e^{-2c} \frac{W(2c)[2c-W(2c)]^{d-1}}{(d-1)!} \\
P(d,m=+1) = \tilde{P}_d (z>0) Po_{2c}(d) &=&
    e^{-2c} \frac{[2c+(d-1)W(2c)][2c-W(2c)]^{d-1}}{(d-1)!}\nonumber
\end{eqnarray}
The results for $c=1$ are displayed in Fig. \ref{figConn} along with
numerical data for $N=17,35,70$. Please note that the numerical
results seem to converge towards the analytical one, thus showing an
excellent coincidence of both approaches. The curves are easily
understood: a vertex with degree 0 has no neighbors. Therefore, it
does not appear in any optimum cover and we obtain $P(0,m=-1)=1$,
$P(0,m>-1)=0$. With increasing degree the probability that a vertex is
covered increases, thus the contribution of $P(k,-1<m<+1)$ to
$\mbox{Po}_{2c}(d)$ increases as well. For large degrees it is very
probable that a vertex belongs to all VCs but even a finite fraction
of vertices with $m=-1$ remains.

This behavior can also be studied by evaluating the average
magnetization $m(d)$ as a function of the degree. Here the analytical
solution gives only lower and upper bounds since we are not able to
precisely calculate the magnetization of the non-backbone spins:
\begin{equation}
    2\left(1+(d-1)\frac{W(2c)}{2c}\right)
    \left(1-\frac{W(2c)}{2c}\right)^{d-1}-1 < m(d) <
    1-2 \left(1-\frac{W(2c)}{2c}\right)^d
\end{equation}
Results are displayed in Fig. \ref{figMagConn}: with increasing size
$N$ of the graphs the numerical data approach the region inside the
bounds. The magnetization turns out to be a monotonously increasing
function of the vertex degree, as expected from the results for
$P(k,m)$. These results justify {\it a posteriori} the application of
the heuristic within the algorithm: vertices having a large degree are
at first included into the cover set.

\subsubsection{Approximating the VC entropy}\label{sec:s_vc}

It is also interesting to go away from the threshold into the
coverable phase, $x>x_c(c)$, and to ask for the number of VCs which is
given by the cover entropy (\ref{s_e0}). As the saddle point equations
for $P(h)$ are to hard to be solved directly, we have used a simple
variational ansatz.  For doing this, we plug a set of simple test
functions into the free energy (\ref{betaf}) and optimize with respect
to these, cf. \cite{BiMoWe} for an application in SAT.  The simplest 
Ansatz is provided by taking a Gaussian distribution,
\begin{equation}
  \label{gauss}
  P^{(var)}(h) = \frac{1}{\sqrt{2\pi\Delta}T} \exp\left\{
         -\frac{(h-Tz_0)^2}{2\Delta T^2} \right\}
\end{equation} 
which includes only two free parameters.
Note that the resulting fields $h$ have already the linear scaling
with temperature $T$ which is needed for the limit $T\to 0$ in the
coverable phase. Using the rescaled variable $z=h/(T\sqrt{\Delta})$, 
we get the following variational expression for the VC entropy:
\begin{eqnarray}
  \label{entropy_gauss}
  s_{VC}^{(var)}(x,c) &=& \int Dz \frac{3-z(z-2z_0/\sqrt{\Delta})}{2}
   \log [2 \cosh( \sqrt{\Delta} z + z_0 )] \nonumber\\
&& +c\int Dz_1 \int Dz_2 \log\left[ 1-\frac{\exp\{-\sqrt{\Delta}
    (z_1+z_2)-2z_0\}}{4 \cosh( \sqrt{\Delta}z_1 + z_0) 
   \cosh(\sqrt{\Delta}z_2 + z_0 )}\right]
\end{eqnarray}
$Dz$ denotes the normal Gaussian measure $dz\ e^{-z^2/2}/\sqrt{2\pi}$.
This expression has to be optimized with respect to the parameters
$\Delta$ and $z_0$ which fulfill the additional constraint
\begin{equation}
  \label{constraint_gauss}
  \int_{-\infty}^\infty Dz \tanh( \sqrt{\Delta} z + z_0 ) = 2x-1\ .
\end{equation}
Fig. \ref{figEntropy} compares the resulting entropy with numerical
enumerations of all VCs for graphs with $c=1.0$ as a function of
$x$. Because of the large numerical effort, only graphs with $N\le 50$
were considered. Deep inside the coverable region, the value of
$s_{VC}^{(var)}$ appears to be a very good approximation, as the
numerical values approach it with increasing graph sizes $N$. Near the
threshold the Gaussian ansatz (\ref{gauss}) starts to fail as it
includes only one scale for the fields and thus is not able to reflect
the partial freezing into backbone and non-backbone spins.  Comparable
results were also obtained for other values of $c$.


\section{Conclusions and outlook}
In this paper the vertex-cover problem on random graphs with a finite
average vertex degree was studied. The problem was investigated using
several methods. Numerical calculations with an exact branch-and-bound
algorithm were performed. The coverability of a graph shows a sharp
transition in the cardinality $xN$ of vertex covers at the threshold
$x_c(c)$. There are almost surely no VCs with $x<x_c(c)$, whereas they
exist almost surely for $x>x_c(c)$. This transitions is related to a
jump in the median complexity of the algorithm, and in the size of the
backbone as well.

A cluster expansion for non-percolated graphs gives very precise
estimates of threshold and backbone for small $c$. Two approaches
coming from the statistical physics of disordered systems were applied
to the VC problem.  The annealed approximation reproduces a known
graph-theoretical lower bound. A more sophisticated method is given by
the replica ansatz, which allows to derive analytical expression for
the threshold $x_c(c)$ and the backbone $b_c(c)$ for average vertex
degrees less than the Eulerian constant $e$, where also the agreement
with numerical data is excellent. These expressions are conjectured to
be exact.  Beyond the average connectivity $2c=e$, the replica
symmetric ansatz fails to produce valuable results, and more
complicated methods including replica symmetry breaking should be
applied in future.

We have also given a variational approximation for the vertex cover
entropy, {\it i.e.} the logarithm of the number of VCs of given
cardinality.  Whereas this approximation was rather precise far above
the covering threshold, the latter can be described only by going
beyond a simple Gaussian approximation. The behavior for 
$x\neq x_c(c)$ deserves  further investigation.

It would also be interesting to consider different graph ensemble,
{\it e.g.} graphs of constant vertex degree or graphs having
locally non-tree-like structures.

\section{Acknowledgements}

The authors are deeply indebted to R. Monasson and R. Zecchina for 
many fruitful discussions on the field of phase transitions in
combinatorial problems. Financial support was provided by the DFG 
({\em Deutsche Forschungsgemeinschaft}) under grant Zi209/6-1.


\appendix

\section{Calculation of the annealed bound}\label{app:annealed}

In this appendix we calculate the annealed bound for the covering
threshold. As stated in (\ref{ann}), it follows from the average
of the partition function over the random graph ensemble. Here
we use the second formulation, see \ref{sec:rg}, where edges are
drawn with probability $2c/N$:
\begin{eqnarray}
  \label{a:ann}
 \overline{Z(T,x|\{J_{i,j}\})} &=& \sum_{{\cal{C}}_x(\{S_i\})}
 \overline{\exp\{-H(\{S_i\},\{J_{i,j}\})/T\}}\nonumber\\
&=& \sum_{{\cal{C}}_x(\{S_i\})} \prod_{1\leq i<j\leq N}
 \overline{\exp\{-J_{i,j}\delta_{S_i,-1}\delta_{S_j,-1}/T\}}
 \nonumber\\
&=& \sum_{{\cal{C}}_x(\{S_i\})} \prod_{1\leq i<j\leq N}
 \left[ 1-\frac{2c}{N}+\frac{2c}{N} 
 \exp\{-\delta_{S_i,-1}\delta_{S_j,-1}/T\} \right] \nonumber\\
&=& \sum_{{\cal{C}}_x(\{S_i\})} \exp\left\{ -cN +\frac{c}{N}
 \sum_{i,j=1}^N \exp\{-\delta_{S_i,-1}\delta_{S_j,-1}/T\} +o(N)
 \right\}\nonumber\\
&=& {N \choose xN} \exp\left\{ cN(1-x)^2 (e^{-T^{-1}}-1) +o(N)
 \right\}\nonumber\\
&=& \exp\left\{N\left[ -x\log(x)-(1-x)\log(1-x)+
 c(1-x)^2 (e^{-T^{-1}}-1)\right]+o(N)\right\}\nonumber
\end{eqnarray}
where the last expression was obtained using Stirling´s formula.
This gives the annealed entropy from section \ref{sec:annealed}
in the limit $T\to 0$.

\section{Calculation of the free energy}\label{app:replica}

The main problem in calculating the free-energy density
consists in the average of the logarithm of the partition
function over the ensemble of random graphs. The replica trick
is based on the simple equality
\begin{equation}
  \label{replica}
  \log Z = \lim_{n\to 0} \frac{Z^n -1}{n}
\end{equation}
which is valid for positive real $Z$. It allows to calculate 
the average of $Z^n$. In principle, this problem is not easier than
before. But the trick used in statistical physics is the following:
We calculate $\overline{Z^n}$ at first for positive integer $n$,
and try to obtain some analytical continuation at the end. The 
$n$-fold power can be understood in terms of $n$ identical copies
$\{S_i^a\}, \ a=1,...,n,$ of the original system. Every of these 
copies has the same Hamiltonian (\ref{hamiltonian}), including 
identical edges $J_{i,j}$,
and fulfills the same magnetization constraint (\ref{constraint}).
The average over random graphs is calculated analogously to the last
appendix, cf. section (\ref{sec:f}) for the notations,
\begin{eqnarray}
  \label{Zn}
Z_n(T,x,c) &:=& \overline{Z^n(T,x|\{J_{i,j}\})} \nonumber\\
&=& \sum_{{\cal{C}}_x(\{S_i^a\})}\overline{ \exp\left\{ -T^{-1} 
    \sum_{i<j} J_{i,j} \sum_{a=1}^n \delta_{S_i^a,-1}  
    \delta_{S_j^a,-1} \right\}} \\
&=& \sum_{{\cal{C}}_x(\{S_i^a\})} \exp\left\{ -cN + \frac{2c}{N}
    \sum_{i<j} \exp\left\{ -T^{-1}\sum_{a=1}^n \delta_{S_i^a,-1}  
    \delta_{S_j^a,-1} \right\} +o(N) \right\} \nonumber
\end{eqnarray}
This can be simplified by introducing the $2^n$ order parameters
which are enumerated by $\vec{\sigma}\in \{+1,-1\}^{n}$:
\begin{equation}
  \label{c_sigma}
  c(\vec{\sigma}) = \frac{1}{N} \sum_{i=1}^N \prod_{a=1}^N
  \delta_{\sigma^a,S_i^a}\ .
\end{equation}
$c(\vec{\sigma})$ measures the fraction of vertices $i$ having 
the replicated spin $(S_i^1,...,S_i^n)=\vec{\sigma}$. We find
\begin{eqnarray}
  \label{Zn2}
Z_n(T,x,c) &=& \int_0^1 \prod_{\vec{\sigma}}\,' dc(\vec{\sigma})
    \frac{N!}{\prod_{\vec{\sigma}} [c(\vec{\sigma})N]! }
    \times \\
 &&\times   \exp\left\{ -cN +cN \sum_{\vec{\sigma},\vec{\tau}} 
    c(\vec{\sigma})c(\vec{\tau}) \exp\left\{ -T^{-1}\sum_{a=1}^n 
    \delta_{\sigma^a,-1}\delta_{\tau^a,-1} \right\}+o(N) \right\}
    \nonumber
\end{eqnarray}    
The integration is over all $c(\vec{\sigma})$ which
are normalized, $\sum_{\vec{\sigma}} c(\vec{\sigma}) =1$, and
fulfill the magnetization constraint, 
$\sum_{\vec{\sigma}} c(\vec{\sigma}) \sigma^a = 2x-1$ for all
$a=1,...,n$. Using Sterlings formula we finally find
\begin{eqnarray}
  \label{Zn3}
Z_n(T,x,c) &=& \int_0^1 \prod_{\vec{\sigma}}\,' dc(\vec{\sigma})
   \exp\left\{ N \left[-c-\sum_{\vec{\sigma}} c(\vec{\sigma})
   \log c(\vec{\sigma}) \right.\right. \nonumber\\
&& \left.\left. +c \sum_{\vec{\sigma},\vec{\tau}} 
    c(\vec{\sigma})c(\vec{\tau}) \exp\left\{ -T^{-1}\sum_{a=1}^n 
    \delta_{\sigma^a,-1}\delta_{\tau^a,-1} \right\} +o(N)
   \right]\right\} \nonumber\\
&=& \exp\left\{ N g_n\left[c_0(\vec{\sigma})\right] + o(N)\right\} 
\end{eqnarray} 
The dominant term of $O(N)$ in the exponent is given by the saddle
point $c_0(\vec{\sigma})$, 
\begin{equation}
   \label{app:saddle}
   \log c_0(\vec{\sigma}) = \lambda_1 + \lambda_2 \sum_a \sigma^a 
   + 2c \sum_{\vec{\tau}}
   c_0(\vec{\tau}) \exp\left\{ -T^{-1}\sum_{a=1}^n 
   \delta_{\sigma^a,-1}\delta_{\tau^a,-1} \right\}    
\end{equation}
where $\lambda_1$ is a Lagrange parameter for  the 
normalization of $c(\vec{\sigma})$, and  $\lambda_2$ a second
one for the magnetization constraint. 

The problem which remains is the continuation to real $n$. 
We have to introduce some ansatz on the structure
of $c_0(\vec{\sigma})$. The simplest one is based on the 
observation, that $Z_n(T,x,c)$ is by definition invariant
under permutations of the $n$ replicas which were introduced as 
being identical. We therefore assume this symmetry also for the
order parameter $c_0(\vec{\sigma})$ which consequently depends
only on $s=\sum_a \sigma^a$. We may express it by a generating
function,
\begin{equation}
c_0(\sigma) = \int_{-\infty}^\infty dh \ P(h) \frac{e^{T^{-1}hs}
      }{ (2\cosh T^{-1}h)^n },
\end{equation}
which is normalized whenever $P(h)$ is normalized, 
$\int_{-\infty}^\infty dh \ P(h)=1$. The magnetization condition
now reads $\int_{-\infty}^\infty dh \ P(h) \ \tanh T^{-1}h=2x-1$.

Plugging this replica symmetric ansatz into 
$g_n [c_0(\vec{\sigma}) ]$, we get (\ref{betaf}) by some 
straight-forward algebra from
\begin{equation}
   f(T,x,c) = -T \lim_{n\to 0} \frac{1}{n} 
          g_n [c_0(\vec{\sigma}) ]\ . 
\end{equation}
Also the saddle point equation (\ref{saddle}) for $P(h)$ can be 
easily calculated from (\ref{app:saddle}).

\section{The saddle point equation at the threshold}\label{app:xc}

In order to calculate the saddle point equation at the threshold,
we take the first procedure proposed in section \ref{sec:xc},
{\it i.e.} we approach the threshold from above, using the scaling
$h=T H_{cov} z$ with some random variable $z$ drawn from the
distribution $\tilde{P}(z)$. In the limit $T\to 0$, 
(\ref{saddle}) slightly simplifies ($\lambda=H_{cov}\mu$):
\begin{equation}
  \int dz\ \tilde{P}(z)\ e^{H_{cov} zs}=\exp\left\{-2c-H_{cov}\mu s 
   +2c \int dz\ \tilde{P}(z)\ \left[ \frac{1}{1+e^{-2H_{cov} z}}
   \right]^{-\frac{s}{2}}  \right\}\ .
\end{equation}
If we approach the threshold, $H_{cov}$ is diverging. In order to
obtain a reasonable limit, we have to keep $t=H_{cov}s$ finite
in this limit:
\begin{eqnarray}
  \int dz\ \tilde{P}(z)\ e^{zt}&=&\exp\left\{-2c-\mu t 
   +2c \int dz\ \tilde{P}(z)\ \lim_{H_{cov}\to\infty}\left[ 
   \frac{1}{1+e^{-2H_{cov} z}}
   \right]^{-\frac{t}{2H_{cov}}}  \right\}\nonumber\\
 &=& \exp\left\{-2c-\mu t 
   +2c \int_{-0}^\infty dz\ \tilde{P}(z) 
   +2c \int_{-\infty}^{-0} dz\ \tilde{P}(z) e^{-tz}\right\}\ .
\end{eqnarray}
Developing the exponential for the last two terms, we find
the desired equation.

\newcommand{\captionPEX}
{Probability $P_{cov}(x)$  that a cover exists for a random realization
 ($c=1.0$) as a function of
  the fraction $x$ of covered vertices. The result is shown for three
  different system sizes $N=25,50,100$ 
(averaged for $10^4$ - $10^3$ samples). Lines are guides to the eyes only.
In the left part, where the
  $P_{cov}$ is zero, the energy $e$ (see text) is displayed. The inset 
  enlarges the result for the energy in the region $0.3\le x \le 0.5$.}

\newcommand{\captionTimeX}
{Time complexity of the vertex cover: 
Median number of nodes visited in the backtracking tree as a
  function of the fraction $x$ of covered vertices for graph sizes
  $N=20,25,30,35,40$ ($c=1.0$). 
The inset shows the  region below the threshold with logarithmic
scale, including also data for $N=45,50$. The fact that in this 
representation the lines are equidistant
demonstrates that the time complexity grows exponentially with $N$.}

\newcommand{\captionTimeCovUncov}
{ 
Median number of nodes visited in the backtracking tree as a
  function of the fraction $x$ of covered vertices, 
displayed separately for the cases of coverable and uncoverable
graphs ($N=30, c=1.0$). Additionally, a scatter plot of the
number of nodes for 100 realizations is presented: for each run a dot
is included in the figure. 
}

\newcommand{\captionBX}
{The fractional size $b(x)$ of the backbone as a function of the relative
  cardinality $x$ of the vertex cover. The results are for the case $c=1.0$
  and for the system sizes $N=25$, 35, 50, 70, and 100. The inset shows results
  for $N=50$. There the fractional backbone sizes are displayed either
  for the subset of graphs which are coverable with $xN$ vertices (cov)
  or for uncoverable graphs (uncov). The total function $b(x)$ is
  almost the minimum of both curves.}

\newcommand{\captionEntropy}
{Entropy of the configurations as a function of the relative cardinality $x$
  of the vertex cover. The symbols represent results from the
  numerical enumerations, for different graph sizes $N$. The solid
  line displays the result from the Gaussian variational approximation.}

\newcommand{\captionXCC}
{Phase diagram: critical fraction $x_c$ of covered vertices as a
function of the edge density $c$. For $x>x_c$, almost all graphs have
covers with $xN$ vertices, while they have almost surely no cover for
$x<x_c$.  The solid line shows the analytic result. The circles
represent the results of the numerical simulations. Error bars are
much smaller than symbol sizes. The upper bound of Harant is given by
the dashed line, the bounds of Gazmuri by the dash-dotted lines.  The
vertical line is at $c=e/2$. Inset: All numerical values were
calculated from finite-size scaling fits of $x_c(N,c)$ using functions
$x_c(N)=x_c+aN^{-b}$.  We show the data for $c=1.0$ as an example.}

\newcommand{\captionBCC}
{The
backbone size $b_c$ at the critical point as a function of $c$.  The
solid line shows the analytic result. The numerical results are
represented by the error bars. They were obtained from finite-size
scaling fits similar to the calculation for $x_c(c)$.
The vertical line is at $c=e/2$. }

\newcommand{\captionConn}
{Distribution of degrees $d$ at the threshold ($c=1.0$). 
We show the total distribution of the degrees, determined by the 
ensemble of random graphs, as well as results describing the minimal
vertex covers. The total distribution is divided into three 
contributions arising from the vertices which either are not in the 
backbone (magnetization $-1<m<1$) or which are in the backbone and 
have magnetizations $m=1$ or $m=-1$. Analytical predictions are 
represented by the lines (which are guides to the eyes only, 
connecting the results for integer arguments), while the numerical 
results for $N=17,35,70$ are displayed using the symbols.}

\newcommand{\captionMagConn}
{The average magnetization of a vertex at the threshold as a function 
of its degree $d$. The lower and upper bounds obtained from the
analytical calculation in the $N\to\infty$ limes are shown by the
lines. The symbols display the numerical results for $N=17,35,70$.}


\begin{figure}[htb]
\begin{center}
\myscalebox{\includegraphics{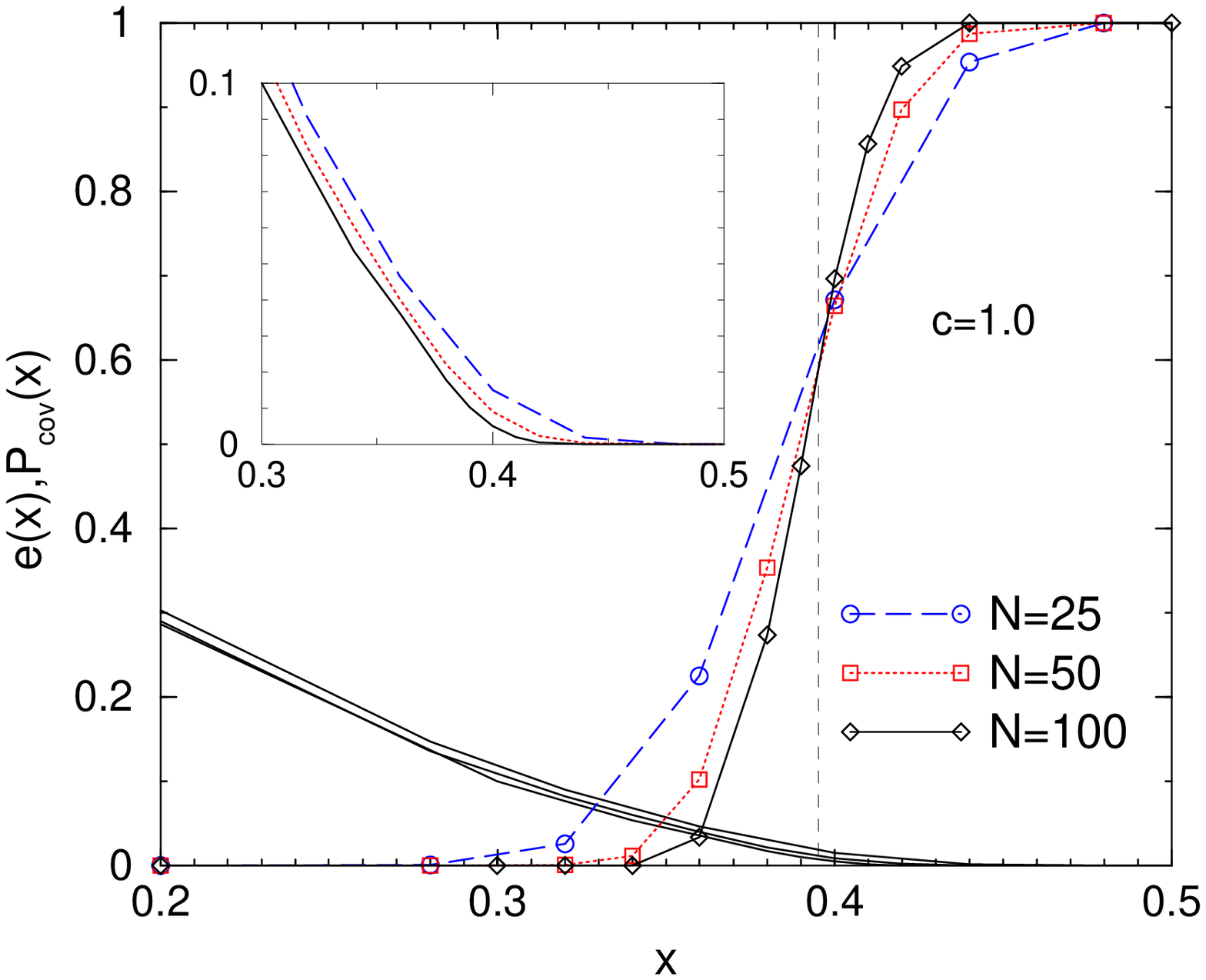}}
\end{center}
\caption{\captionPEX}
\label{figPEX}
\end{figure}

\begin{figure}[htb]
\begin{center}
\myscalebox{\includegraphics{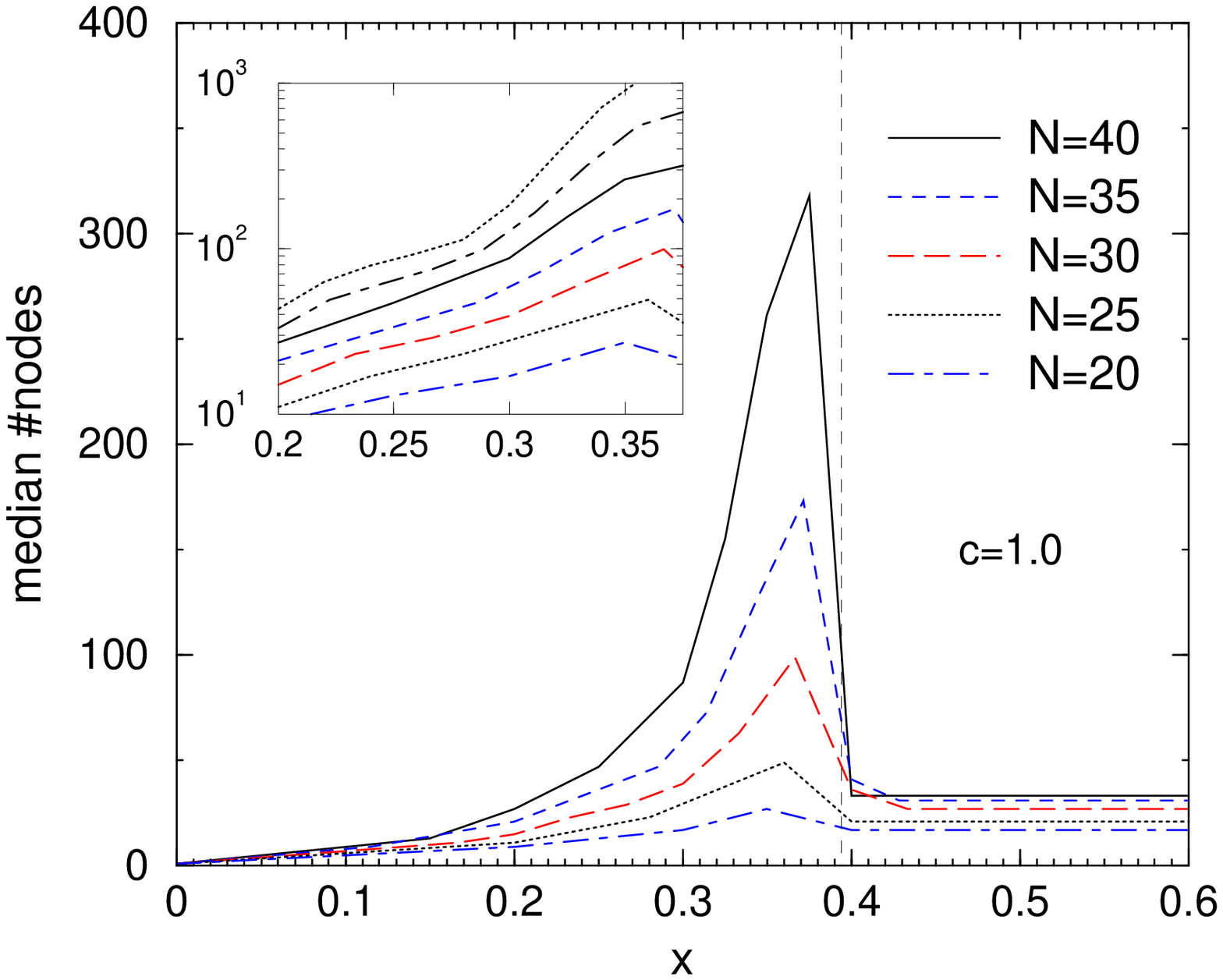}}
\end{center}
\caption{\captionTimeX}
\label{figTimeX}
\end{figure}

\begin{figure}[htb]
\begin{center}
\myscalebox{\includegraphics{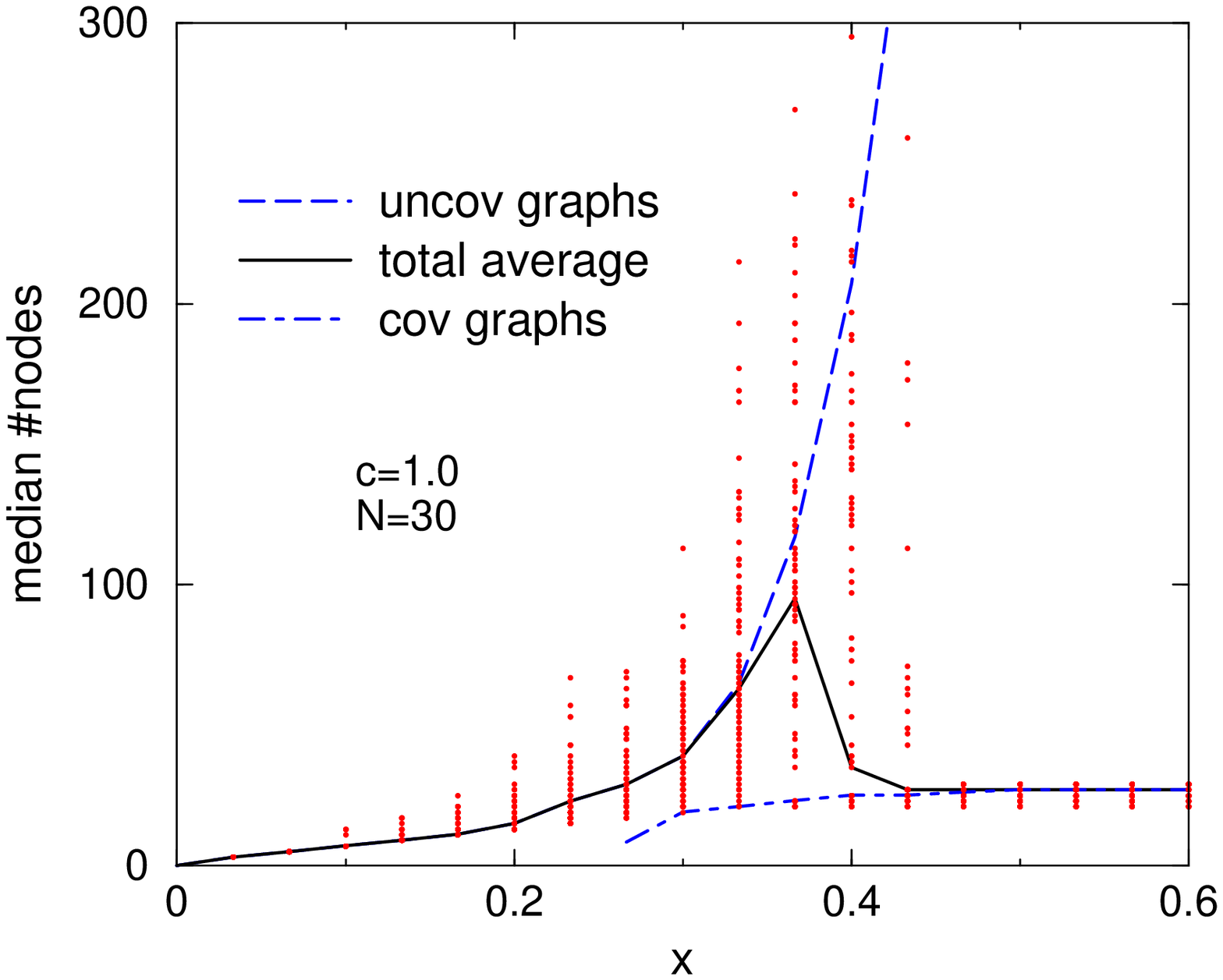}}
\end{center}
\caption{\captionTimeCovUncov}
\label{figTimeCovUncov}
\end{figure}

\begin{figure}[htb]
\begin{center}
\myscalebox{\includegraphics{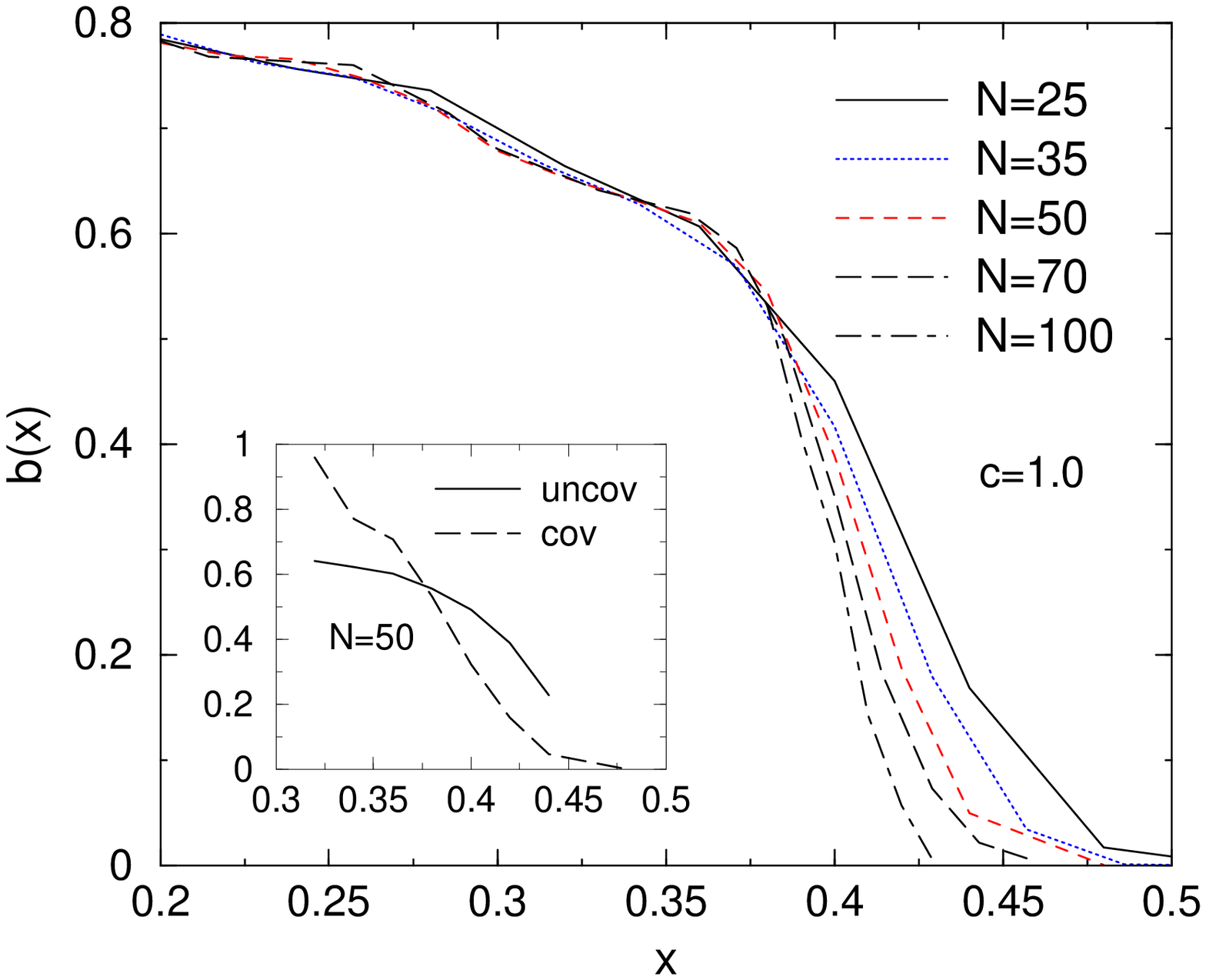}}
\end{center}
\caption{\captionBX}
\label{figBX}
\end{figure}

\begin{figure}[htb]
\begin{center}
\myscalebox{\includegraphics{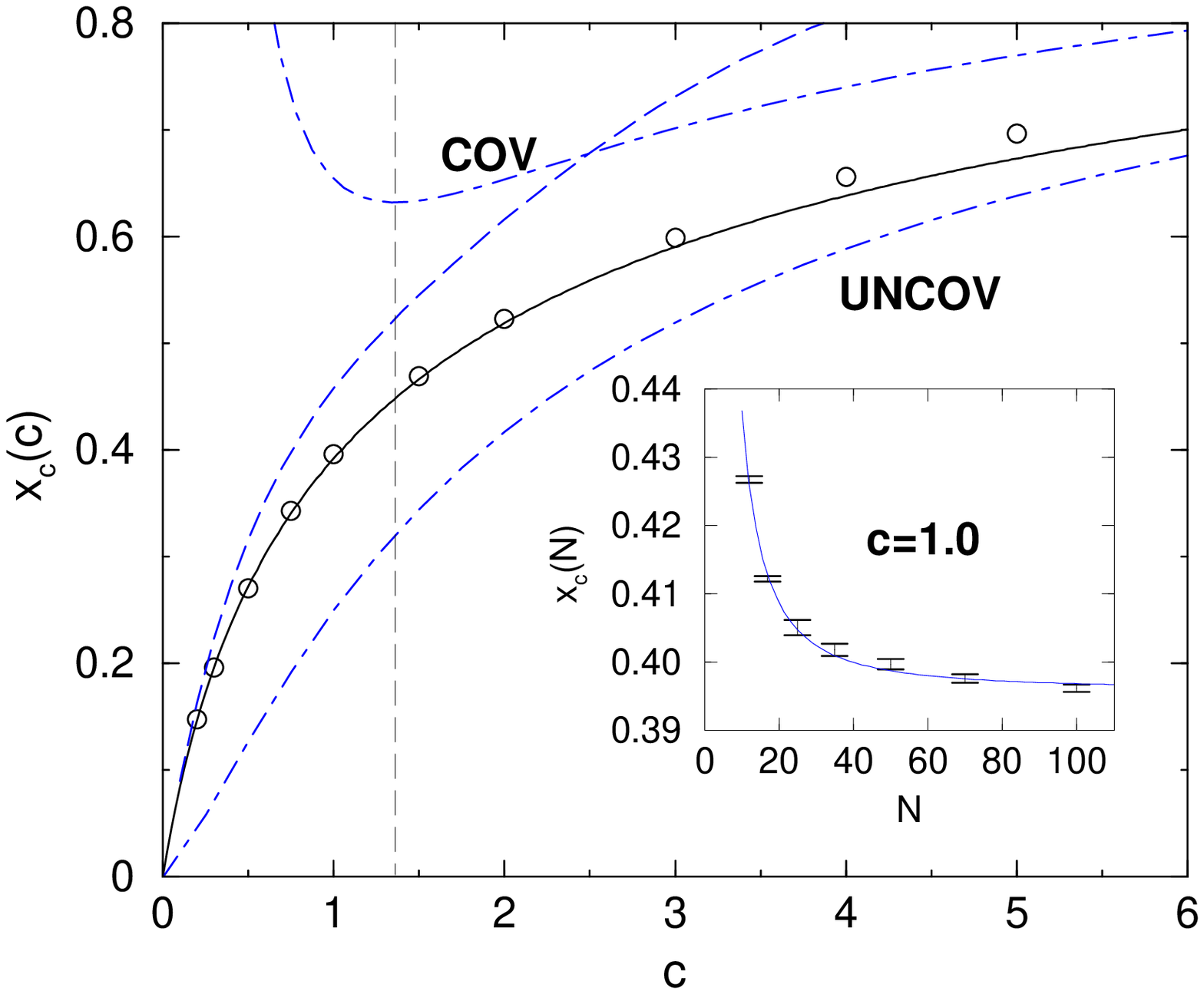}}
\end{center}
\caption{\captionXCC}
\label{figXCC}
\end{figure}

\begin{figure}[htb]
\begin{center}
\myscalebox{\includegraphics{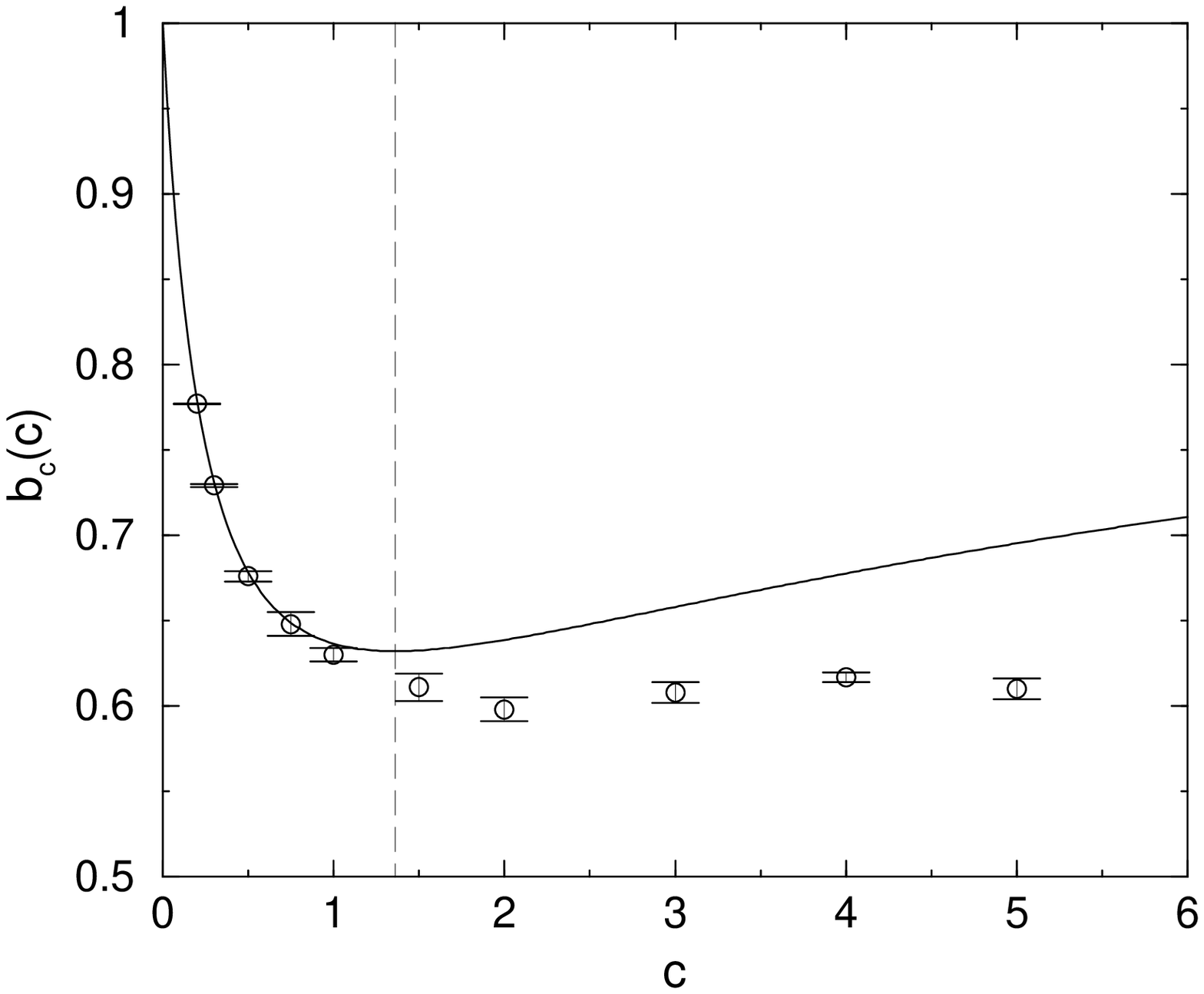}}
\end{center}
\caption{\captionBCC}
\label{figBCC}
\end{figure}

\begin{figure}[htb]
\begin{center}
\myscalebox{\includegraphics{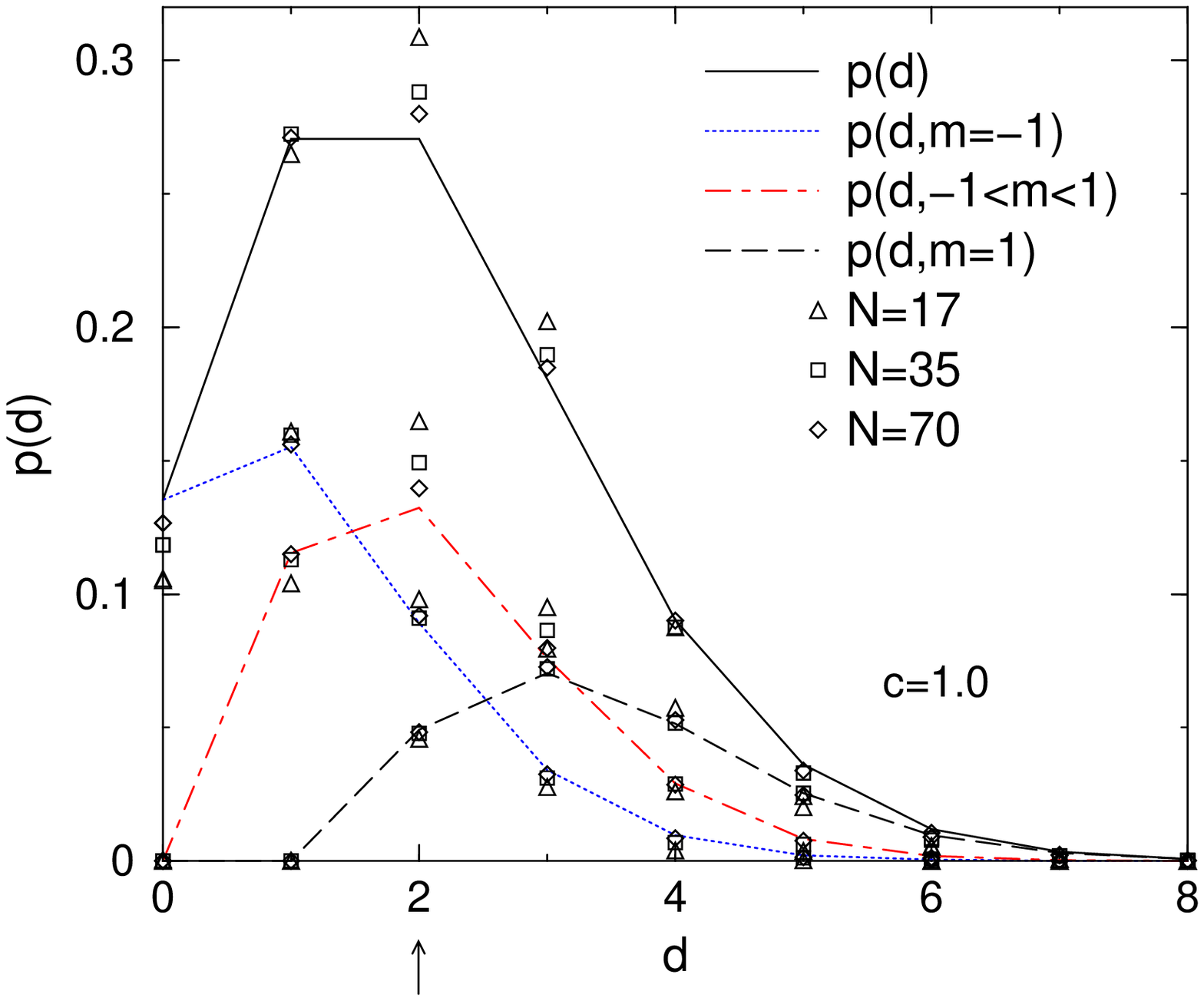}}
\end{center}
\caption{\captionConn}
\label{figConn}
\end{figure}

\begin{figure}[htb]
\begin{center}
\myscalebox{\includegraphics{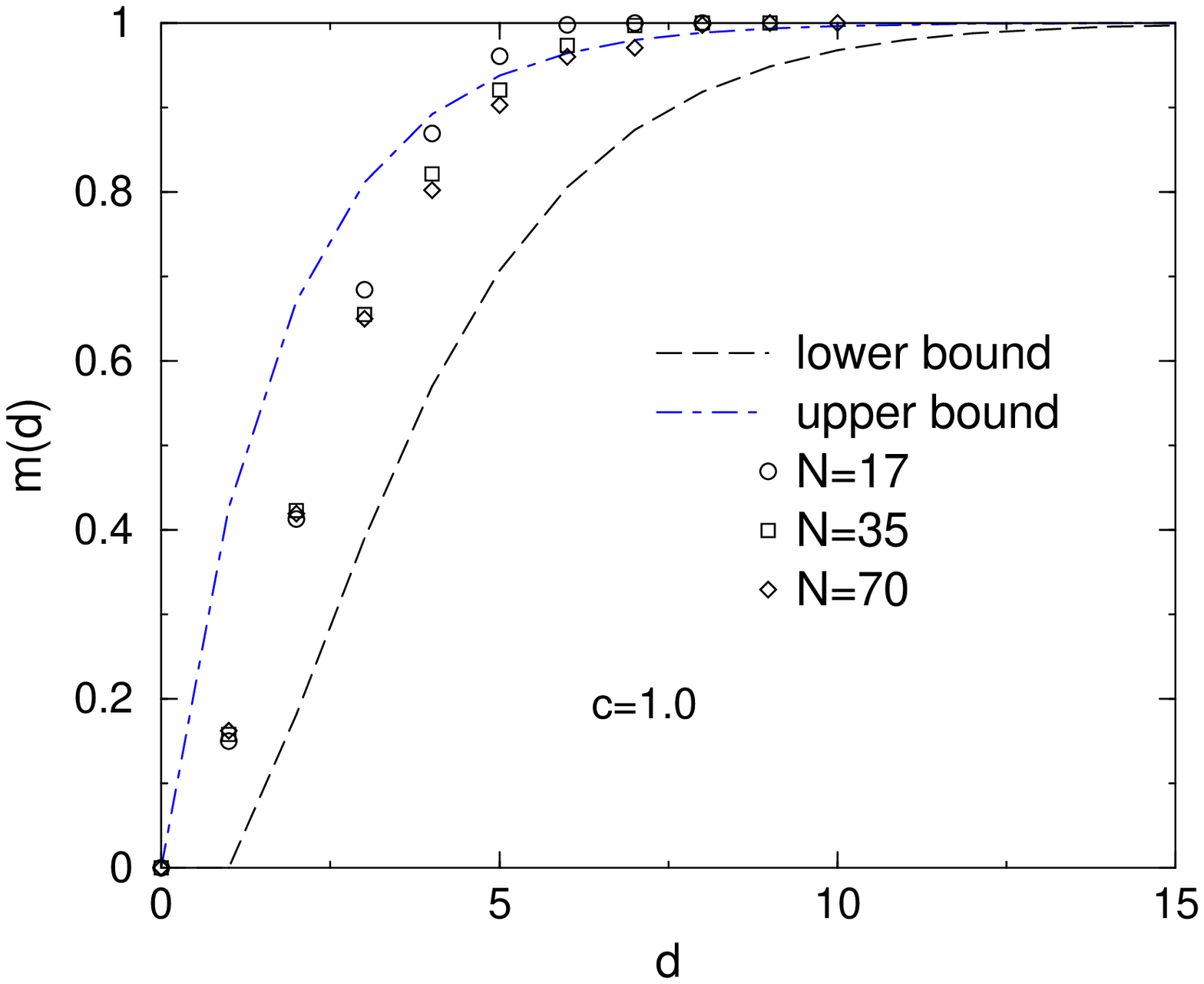}}
\end{center}
\caption{\captionMagConn}
\label{figMagConn}
\end{figure}

\begin{figure}[htb]
\begin{center}
\myscalebox{\includegraphics{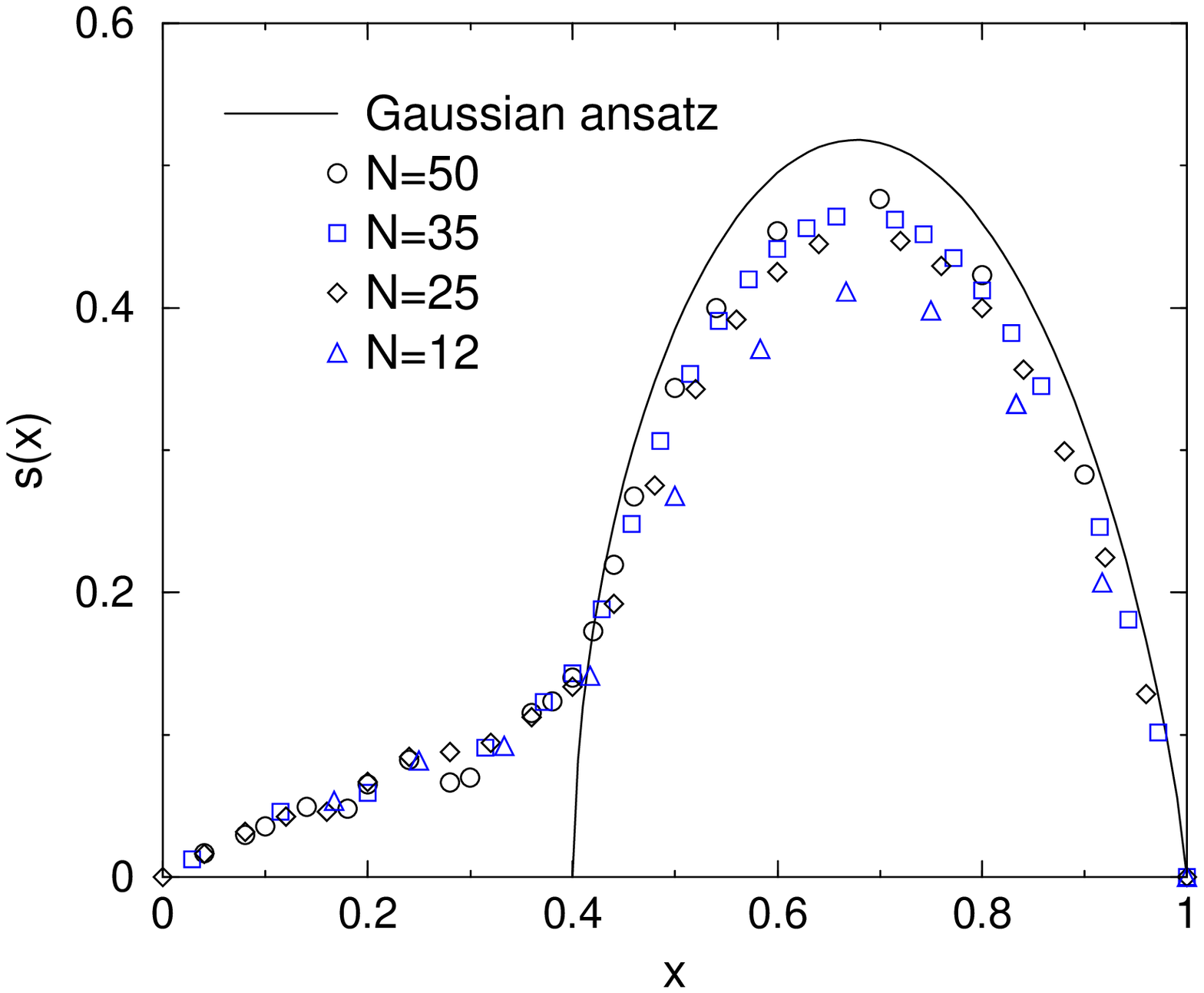}}
\end{center}
\caption{\captionEntropy}
\label{figEntropy}
\end{figure}


\end{document}